\documentclass[twocolumn,preprint]{aastex63}

\usepackage{amsmath,amstext}
\usepackage[T1]{fontenc}
\usepackage{apjfonts} 
\usepackage{graphicx}
\usepackage{natbib}
\usepackage{longtable}

\usepackage{microtype,longtable,verbatim,float,booktabs,graphicx}

\newcommand{\Ha}{\hbox{{\rm H}$\alpha$}}
\newcommand{\Hb}{\hbox{{\rm H}$\beta$}}

\newcommand{\HavNII}{\hbox{({\rm H}$\alpha$+{\rm [N}\kern 0.1em{\sc II}{\rm ]})}}

\newcommand{\HnFeV}{\hbox{{\rm H}9+{\rm [Fe}\kern 0.1em{\sc V}{\rm ]}}}

\newcommand{\HeH}{\hbox{{\rm H}8+{\rm He}\kern 0.1em{\sc I}{\rm }}}

\newcommand{\HeII}{\hbox{{\rm He}\kern 0.1em{\sc II}}}

\newcommand{\HeI}{\hbox{{\rm He}\kern 0.1em{\sc I}}}

\newcommand{\HII}{\hbox{{\rm H}\kern 0.1em{\sc II}}}

\newcommand{\Ms}{M_\star}

\begin{document}

\title{\large \bf MEGA Mass Assembly with JWST: The MIRI EGS Galaxy and AGN Survey}


\author[0000-0001-8534-7502]{Bren E. Backhaus}
\affil{Department of Physics and Astronomy, University of Kansas, Lawrence, KS 66045, USA}

\author[0000-0002-5537-8110]{Allison Kirkpatrick}
\affiliation{Department of Physics and Astronomy, University of Kansas, Lawrence, KS 66045, USA}

\author[0000-0001-8835-7722]{Guang Yang}
\affiliation{Nanjing Institute of Astronomical Optics and Technology, Nanjing 210042, China}

\author[0000-0001-5930-0532]{Gregory Troiani}
\affil{Department of Physics and Astronomy, University of Kansas, Lawrence, KS 66045, USA}

\author[0000-0002-6292-4589]{Kurt Hamblin}
\affiliation{Department of Physics and Astronomy, University of Kansas, Lawrence, KS 66045, USA}

\author[0000-0001-9187-3605]{Jeyhan S. Kartaltepe}
\affiliation{Laboratory for Multiwavelength Astrophysics, School of Physics and Astronomy, Rochester Institute of Technology, 84 Lomb Memorial Drive, Rochester, NY 14623, USA}

\author[0000-0002-8360-3880]{Dale D. Kocevski}
\affiliation{Department of Physics and Astronomy, Colby College, Waterville, ME 04901, USA}

\author[0000-0002-6610-2048]{Anton M. Koekemoer} 
\affiliation{Space Telescope Science Institute, 3700 San Martin Drive, Baltimore, MD 21218, USA} 

\author[0000-0003-3216-7190]{Erini Lambrides}\altaffiliation{NPP Fellow}
\affiliation{NASA-Goddard Space Flight Center, Code 662, Greenbelt, MD, 20771, USA}

\author[0000-0001-7503-8482]{Casey Papovich} 
\affiliation{Department of Physics and Astronomy, Texas A\&M University, College Station, TX, 77843-4242 USA} 
\affiliation{George P.\ and Cynthia Woods Mitchell Institute for Fundamental Physics and Astronomy, Texas A\&M University, College Station, TX, 77843-4242 USA} 

\author[0000-0001-5749-5452]{Kaila Ronayne} 
\affiliation{Department of Physics and Astronomy, Texas A\&M University, College Station, TX, 77843-4242 USA} 
\affiliation{George P.\ and Cynthia Woods Mitchell Institute for Fundamental Physics and Astronomy, Texas A\&M University, College Station, TX, 77843-4242 USA} 


\begin{abstract}

We present the MIRI EGS Galaxy and AGN (MEGA) survey, a four band MIRI survey with 25 pointing in the Extended Groth Strip (EGS) extragalactic field. Three of the pointings utilized only the three reddest bands (F1000W, F1500W, F2100W) while the remainder of the pointings also add a blue filter (F770W). MEGA builds upon the existing observations in the EGS field by providing MIRI imaging for 68.9\% of CEERS NIRCam imaging, filling a cruciality gap in order to understand galaxy evolution by observing the obscured Universe. Here, we present the technical design, data reduction, photometric catalog creation for our the first data release, and science drivers of the MEGA survey. Our data reduction starts with the standard JWST calibration pipeline, but adds additional warm pixel masking and custom background subtraction steps to improve the quality of the final science image. We estimate the image depth of the reduced mosaics and present new galaxy number counts in four MIRI bands.

\end{abstract}

  

\section{Introduction}

Mid-infrared (MIR) observations are able to peer through dust, allowing for a clear view of galaxies within. This allows for accurate measurements of a galaxies properties and study how galaxies form and evolve across time. Previous telescopes did not provide the ability or sample needed to study the obscured galaxies at higher redshifts. Observations to peer past the dust using \textit{Spitzer} is unable to observed the resolved interstellar medium for galaxies $z>1$ due to its low resolution (6'' at 24 $\mu$m). Likewise, the Wide-field Infrared Survey Explorer (WISE) also did not have a high enough resolution (12'' at 22 microns). 
While observations from ALMA have focused on the most massive or lensed galaxies \citep{Hodge2015,Hodge2016}. 
Now, with the Mid-Infrared Instrument \citep[MIRI;][]{Rieke2015,Bouchet2015,Wright2023}
on the James Webb Space Telescope \citep[JWST;][]{Rigby2023,Gardner2023}
we have a high enough resolution (0.67" at 21 $\mu$m) to observed the obscured galaxies up to z$\sim$3. 

\begin{figure*}[tbp]
\centering
\epsscale{1.1}
\plotone{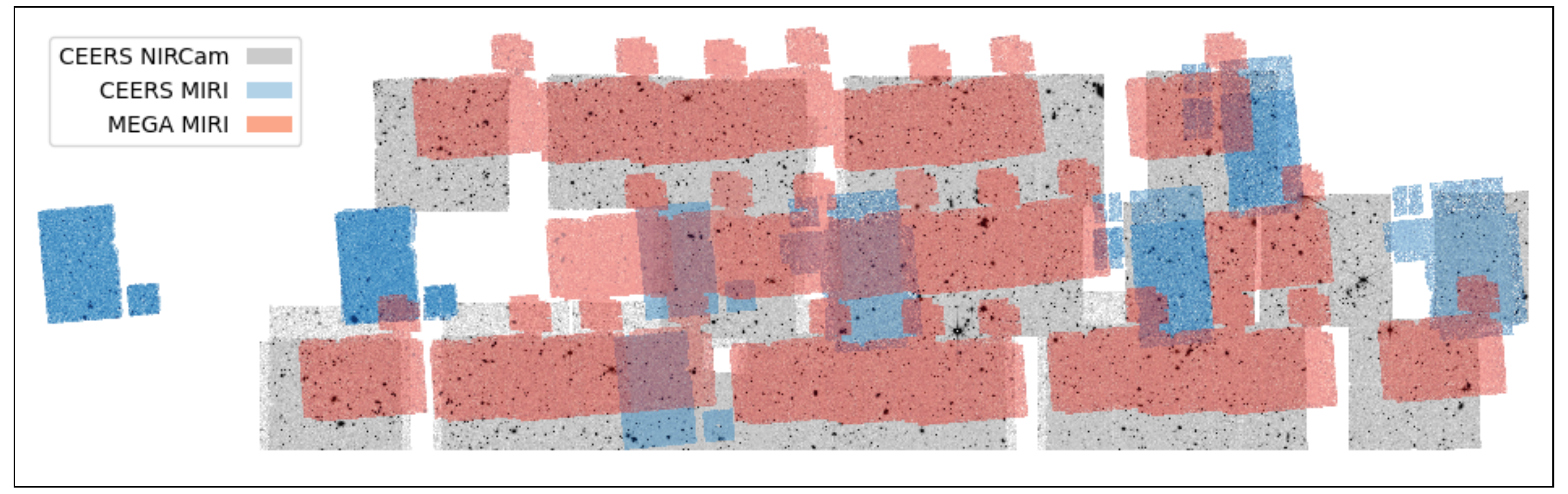}
\caption{MEGA MIRI imaging (red), with CEERS MIRI imaging (blue), and  CEERS NIRCam imaging (gray).
\label{fig:CEERSvMEGA}} 
\end{figure*}

MIRI has given us access to resolve the obscured universe at an important time in galaxy evolution. Cosmic Noon ($z=1-3$) is a key time period to understand galaxy formation and evolution, as 50\% of a galaxies stellar mass is formed. Previous work has shown that at least half of the light from star formation and accretion onto supermassive black holes (SMBHs) is absorbed by dust \citep{mada14,Hickox2018}. While Cosmic Noon has had unobscured star formation measurements from UV/Optical emission lines such as \Ha\ and $\Hb$, MIRI now allows the obscured star formation rate to be measured by polycyclic aromatic hydrocarbon (PAH) emission features at 6.2, 7.7, 11.2, and 12.7 $\mu$m \citep{Ronayne2024,Shipley2016,Spoon2007}. 

There is ample evidence that most or all of massive galaxies host a  supermassive black hole (SMBH) at their centers, the peak of SMBH's accretion rate density occurs at Cosmic Noon \citep{Delvecchio2014,Peca2023}. MIRI allows us to observe the dusty toroidal structure surrounding the accretion disk typically seen in AGN, as this region radiates as a power-law in the mid-IR. Being able to accurately measure the growth of SMBH's and the galaxies star formation rate allow us to build our fundamental understanding of how stellar and black hole mass evolve together. Additionally, these features make the JWST MIRI observations a crucial tool to identifying populations of star-forming galaxies and both obscured and unobscured active galactic nuclei (AGN).

The distribution of sources detected in a given filter as a function of flux, also known as number counts, reveals the evolution of galaxies through cosmic time. Due to number counts requiring only sufficient depth, area, and robust measurements of source fluxes it is subject to fewer sources of calibration bias when comparing to simulations than luminosity or mass functions. The MIR number counts differ significantly from number counts in optical or near-IR filters. \cite{Fazio2004}, found that MIR number counts have an excess at bright magnitudes when compared to the Euclidean world model. \cite{Papovich2004} found Spitzer 24 $\mu$m number counts rise rapidly at near-Euclidean rates down to 5 mJy, however increase with a super-Euclidean rate between 0.4 and 4 mJy, before converging below 0.3 mJy. MIRI now gives access sources fainter than ever observed, allowing further constraints on the faint end models which is the region with the most variance between different cosmological simulations. 
Previous work with JWST/MIRI 7.7, 10, and 15 $\mu$m band number counts show massive improvements in the image depth \citep{kirkpatrick2023}, and have been used to help further constrain galaxy and black hole evolution models \citep{Kim2024}. However, previous work primarily covered small areas of $\sim$5arcmin$^2$ in each field this makes the number counts more sensitive to cosmic variance. The SMILES survey however covered wider area 34 arcmin$^2$ with complete MIR wavelength coverage found that their number counts diverged from predictions from recent semi-analytical models of galaxy formation \cite{Stone2024}.


In this paper, we discuss the MIRI EGS Galaxy and AGN (MEGA) survey MIRI observation, reductions, quality and provide number counts for the four filters used. In Sections \ref{Observations} we describe the observations for our survey. In Section \ref{Data_Reduction} and \ref{Phot_Cat} we discuss the details of the data reductions and how our photometric catalog was created. In Section \ref{Completeness_Num_Count} we assess the quality of our reduced data products. In Section \ref{Science} we provide an overview of the science goals of the MEGA survey. Finally, in Section \ref{Summary} we summarize
our results.

\section{Observation Set Up}\label{Observations}

The MIRI EGS Galaxy and AGN (MEGA; PID: 3794; PI: A. Kirkpatrick) is a 67 hour JWST MIRI imaging program covering 70 arcmin$^2$ of the  Extended Groth Strip (EGS) field. MEGA has 25 MIRI pointing taken in March 2024, using the F770W, F1000W, F1500W, and F2100W filters. Pointings 3, 6, and 10 only use the filters F1000W, F1500W, and F2100W. MEGA uses a 4-point dither pattern optimiazed for extended sources and use the FAST1 readout pattern. Table \ref{tab:filter_info} outlines the exposure set up for each filter, there 5$\sigma$ depth in microJanskys, and the number of galaxies with a 3$\sigma$ detection in each filter. MEGA was designed to coincide with the ERS CEERS NIRCam observations (CEERS; PID: 1345; PI: Steven Finkelstein), which covers six filters F115W, F150W, F200W, F277W, F356W, F410W, and F444W \citep{Finkelstein2025}. In addition to CEERS MEGA builds upon the wealth of observation in the EGS field.  These observations also overlap with the Hubble Space Telescope (HST) imaging F125W, F160W, F606W, and F814W \citep{grog11, koek11}, and the Chandra X-ray observations. Figure \ref{fig:CEERSvMEGA} depicts how the MEGA MIRI observations overlap with CEERS NIRCam and MIRI observations. 68.9\% of MEGA MIRI observations have CEERS NIRCam coverage.


\begin{deluxetable*}{c|c|c|c|c|c|c}[t]
\tablecaption{Observational Set up \label{tab:filter_info}}
\tablenum{1}
\tablecolumns{7}
\tablewidth{0pt}
\tablehead{\colhead{Filter} &\colhead{$N_{groups}$}&\colhead{$N_{int}$} &\colhead{Exp. time [s]}& \colhead{5$\sigma$ limit [$\mu$Jy]} & \colhead{80\% Completeness}& \colhead{$N_{Gal}$}}
\startdata
F770W & 1 & 100 & 1100 & 0.18 & 0.14 &4444\\
    F1000W & 1 & 100 & 1100 & 0.41 & 0.30 &3459\\
    F1500W & 3 & 40 & 1354 & 1.26 & 1.08 &2924\\
    F2100W & 6 & 30 & 2054 & 4.10 & 3.31 &2168\\
\enddata
\caption{Exposure setup and Image depth of MEGA MIRI observations which are measured with an aperture with a radius equal to the FWHM of each filter, see Section 5.1. $N_{gal}$ represents the number of galaxies with a 3$\sigma$ measurement in each filter.}
\end{deluxetable*}

\section{Data Reduction} \label{Data_Reduction}

This section describes the data reduction for MEGA,
which followeds a similar procedure to the reductions done in previous works \citep{Yang2023,Alberts2024,Morrison2023}. Data reduction is done the JWST Calibration Pipeline v1.14.0 with JWST Calibration Reference Data System
(CRDS) version: 11.17.19 and CRDS context 1241. Additional, steps such as identifying and masking warm pixels and custom background subtraction were added into our reduction process. All the MEGA data used in this paper can be found in MAST: \dataset[10.17909/9nab-af49]{http://dx.doi.org/10.17909/9nab-af49}.



\subsection{Stage 1}

Stage 1 starts with the with the raw data "uncal.fits" file from MAST and then preforms detector-level corrections before reducing each four-dimensional ramp data set to a two-dimensional rate map. This stage of the pipeline initializes the data quality (DQ) arrays, corrects for known noise patterns, identifies saturated pixels, flags the flags the first and last group in each iteration as bad, corrects for the reset anomaly effect,  non-linearity corrections are applied, the dark current is subtracted, and finds and flags outlier pixels. For this stage, we adopt the default parameters in the pipeline. Based on \cite{Morrison2023}, which notes several artifacts which are not currently addressed in the pipeline, we will correct for cruciform and row/column striping in between stage 2 and 3.

The output of stage 1 pipeline is uncalibrated count rate ("rate.fits") images. These are fed into the stage 2 JWST Pipeline.

\subsection{Stage 2}

Stage 2 of the JWST Pipeline applies additional corrections and calibrations to the rate.fits file produced in stage 1 to create calibrated science images ("cal.fits" files). The corrections and calibrations done in this Stage included preforming image-from-image subtraction to accomplish subtraction of background signal, associating a WCS object with each exposure, dividing out a flatfield, applying flux calibrations, and resampling each exposure into a distortion-corrected 2D image. This Stage converts the image from counts/s to MJy/sr. For Stage 2 of the JWST Pipeline we adopt the default parameters in the pipeline however we skip the background subtraction step. After the resulting "cal.fits" files are created in this step we then remove warm pixels, apply a custom background subtraction, and correct the astromtry.

\subsubsection{Warm Pixel Masking}

\cite{Alberts2024} found that faint outliers had evaded detection in both the jump detection step in stage 1 and the outlier detection step in stage 3 in there earlier reductions of there data. They found that these faint outliers were traced all the way back to the individual "cal.fits" files produced in stage 2. This work found these faint outliers were the same pixels in each cal.fits file when grouped by filter and time the observation was taken. 

We followed the same masking procedure proposed by \cite{Alberts2024}. This is done by median stacking all "cal.fits" files on a per filter basis, masking the pixels which are already flagged in the DQ mask, then the flagging pixels that are $>10\sigma$ above the median of all pixels of the median-stacked image in the the DQ mask. By stacking all the "cal.fits" files on a per filter basis the real sources within the individual images are masked as we have 88-100 cal.fits files per filter over 25 pointing allowing the stacked image to represent the background of the filter. In this work, we chose to use $>10\sigma$ instead of the proposed $>3\sigma$ in \cite{Alberts2024} because we found that due to large number of dithers (88-100) the lower cutoff would cut off $\sim$21\% of the pixels.
With a $>10\sigma$ cut off we found $0.025\% -0.076\%$ of our pixels were flagged as warm in our four filters. This precentage of pixels flagged is similar to \cite{Alberts2024}, which had $0.03\% -0.18\%$ across 8 filters.


\begin{figure*}[tbp]
\centering
\epsscale{1.1}
\plotone{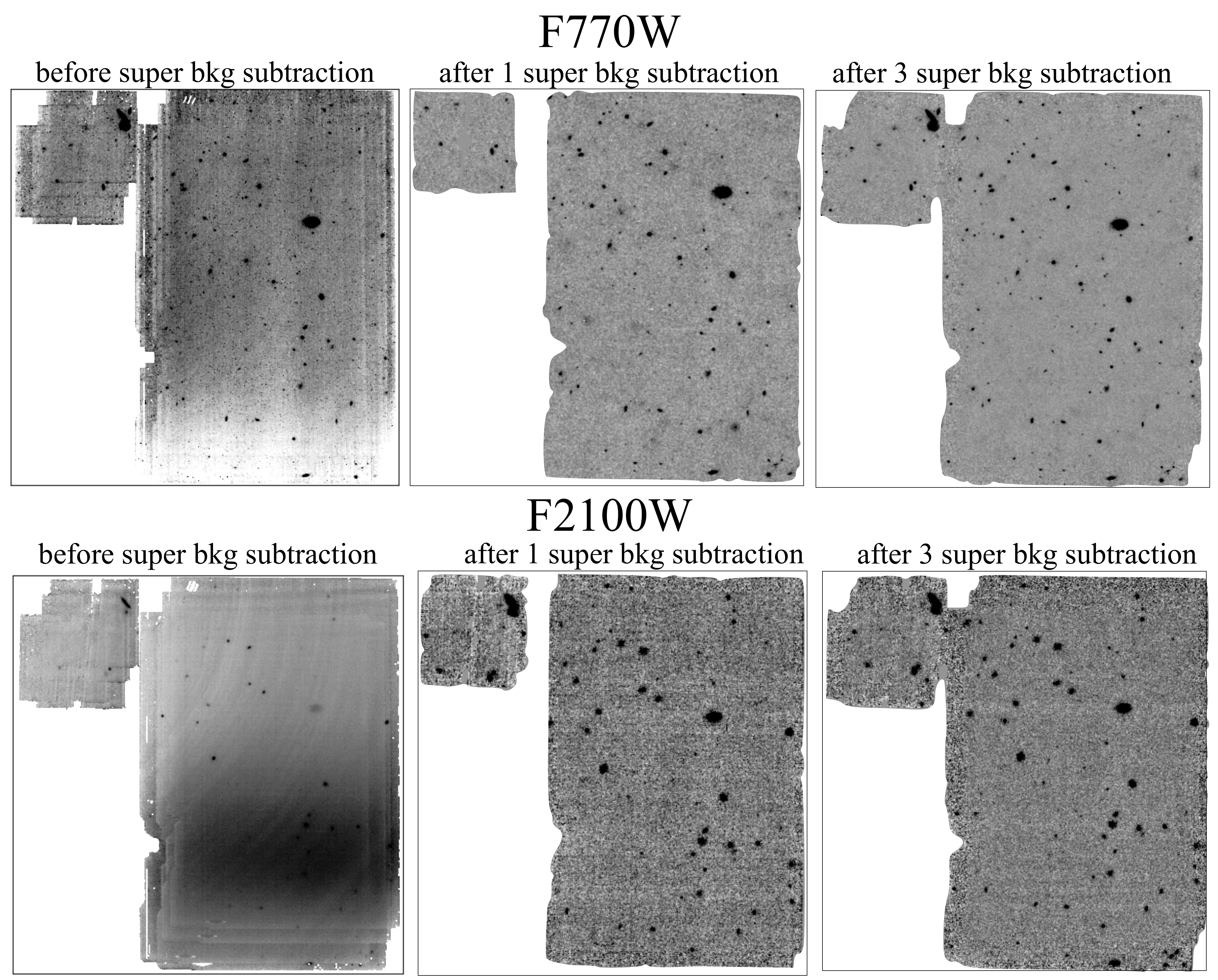}
\caption{ Final mosaic before any background subtraction (left), after one iteration of super background subtraction (middle), and after three iterations of super background subtraction (right) for F770W filter (top) and F2100W filter (bottom). With further iterations of super background subtraction the row and columns striping improves and produces a more uniform background.
\label{fig:reduction}} 
\end{figure*}

\subsubsection{Custom Background Subtraction}

We implement the custom "super background" strategy used in \cite{Alberts2024}, similar to those used in \cite{Yang2023}, \cite{Papovich2023}, and \cite{Perez2024}. This version of background subtraction takes advantage of the large number of dithers (88-100) per filter. The super background strategy starts by putting the resulting warm pixel corrected cal files through default stage 3 pipeline without the background or tweakreg steps to create a mosaic ("i2d.fits" file). These mosaics are then background subtracted using \texttt{Background2D} from \texttt{photutils}. A source mask is then created using \texttt{make$\_$source$\_$mask} from \texttt{photutils}, starting with a high detection threshold of 10$\sigma$. The resulting source mask is then mapped back onto the warm pixel corrected individual cal.fits files using \texttt{reproject$\_$interp}. 
These cal.fits files are then corrected for the vertical/horizontal strip using the method in \cite{Yang2023} which calculates the median of the pixels in each row/column, using a sigma clip of 2$\sigma$ to avoid pixels illuminated by sources then subtract this median from the pixels in the row/column. These files are then background subtracted using \texttt{Background2D}. These resulting cal.fits files are then put back into the stage 3 pipeline with the same settings used previously to create another mosaic. This is done two more times for optimal background subtraction which each iteration using lower source detection thresholds of 6 and 4 respectively.

The final mask of all the sources is then mapped back onto the original un-filtered warm pixel corrected cal.fits files. The median value of the individual masked warm pixel corrected cal.fits files are then subtracted to bring the median background value to zero. A super background for each individual un-filtered warm pixel corrected cal.fits is then created by stacking all the other background subtracted cal.fits files in the filter. This super background is then scaled to the individual cal.fits files and subtracted. These super background subtracted cal.files are then corrected for row/columns striping and removed and remaining background using the \texttt{Background2D}. These corrected super background cal.fits files are what is used for our final Stage 3 mosaic images.

This is done for all pointing except Pointing 18, which uses the stack of the four cal.fits files in each of its filters. We found this pointing's background to be different enough from the median stack to cause an artifical gradient in the dark current and "tree rings". 
While, \cite{Morrison2023} found that and detector effects such as row/column striping and "tree rings that" are generally static for the detector for a given dataset, MEGA appears to be a large enough dataset to have a larger amount of deviation between pointings.
The deviations in Pointing 18 are traced all the way back to the \_uncal.fits data, where the pixels of Pointing 18 are $\sim$50-100 counts lower than the rest of the pointings. We find that this pointing does still have a comparable error mosaic, meaning the final science image of Pointing 18 will have a lower SNR than the rest of the sample.

Figure \ref{fig:reduction} shows the effects of multiple super background subtractions for the final mosaics. A single super background subtraction immediately removes the ripple features while greatly improving the row/column striping and background gradients. However, after three iterations of background subtraction the row/column striping and background gradients continue to improve particular for the bluer filters.



\subsection{Stage 3}

\begin{figure*}[tbp]
\centering
\epsscale{1.1}
\plotone{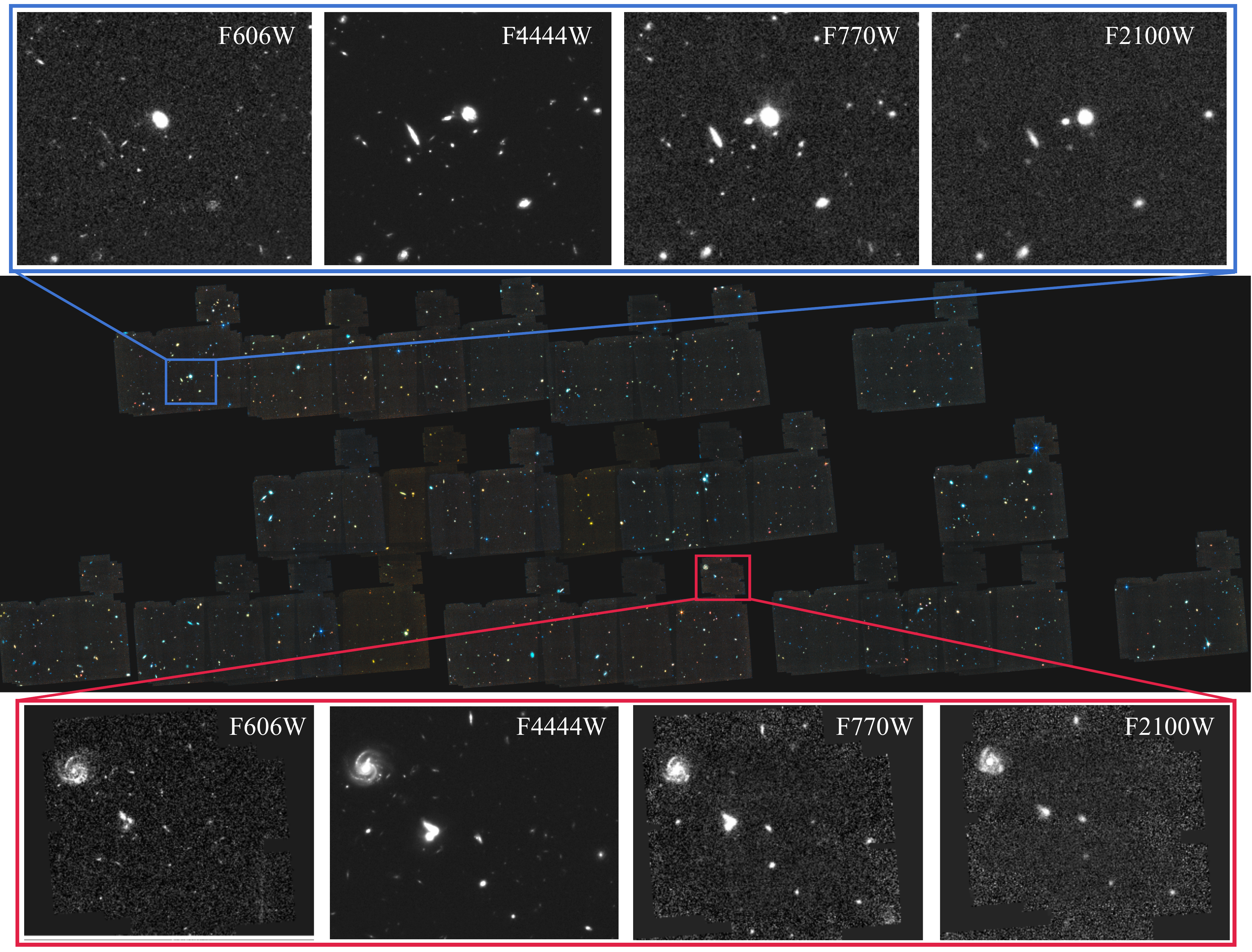}
\caption{RBG image of MEGA Pointings using all four of the MEGA filters. The zoomed-in areas show HST F606W,NIRCam F444W, MIRI F770W, and MIRI F2100W. 
\label{fig:mosaic}} 
\end{figure*}

Stage 3 performs additional corrections for each dither image and then combines all the individual images ("cal.fits" files) and dithers into a single mosaic per filter ("i2d.fits" file). The corrections and steps in stage 3 includes astrometric alignment, compute and matching sky values, outlier detection, and resampling the images onto a common output grid producing  a single undistorted image. Our parameters and modifications for each step in stage 3 are outlined in this section.

The first step known as "tweakreg" corrects the astrometry. The default inputs attempts to align MIRI astrometry to the Gaia DR3
astrometric reference frame (Gaia Collaboration 2022).
However, \cite{Yang2023} found that due to MIRI’s small field of view ($\sim$ 2 arcmin$^2$) there are very few Gaia objects in a MIRI pointing causing the default tweakreg step to fail. To combat this we adopt \cite{Bagley2023} modification for the TweakReg raw code to use the CEERS NIRCam data as the reference frame rather than Gaia. This modified TweakReg step aligns each MIRI
dither with the HST/F160W reference mosaic \cite{koek11} in the EGS field. This modification needs a MIRI source list, so we use SEP to perform source detection for this step. Additionally, this modification minimizes the median absolute error (MAE) due to MAE being less sensitive to outliers. Whereas, the default TweakReg minimizes the rms error between MIRI and the reference. We used the tweakreg parameters in \cite{Yang2023} of tolerance = 0.5, searchrad = 1, and separation =
0.01 and run this modified tweakreg step twice to optimize our result.

For the final resampling step, we drizzle the final mosaic into a WCS frame with the same tangent point as the CEERS HST mosaic in order to match the previous MIRI reductions on the EGS field. The R.A. and decl. are set to 214.825 and 52.825, respectively, using the "crval" parameter. Additionally, we set the "pixel$\_$scale" to be 0.''09 (three times the HST mosaic) and "rotation" to -49.7 degs to match the HST mosaic. These settings create a final MIRI mosaic shown in Figure \ref{fig:mosaic}, which aligns with the HST mosaic and the CEERS MIRI/NIRCam imaging data products. This figure also shows zoomed in regions of the mosaic in the HST F606W, MIRI F707W, and MIRI F2100W filters.

\section{Photometric Catalog} \label{Phot_Cat}


\subsection{Source Detection}\label{Source_Detection}

Our source detection method is similar to that done in \cite{Alberts2024}, which creates a SNR "detection" image using the science and error mosaics. The error mosaics produced by the pipeline is constructed from a quadrature sum of sky, read, and Poisson noise. \cite{Alberts2024} creates this SNR image from the F560W and F770W filters, however our SNR image is the ratio of the inverse-variance weighted stacks of either the F770W or F1000W mosaics. Using the lowest MIRI filter allows the SNR image to have a higher resolution, which will allow for better deblending of sources. This source detection method is based on algorithms created to detect low-surface brightness features in deep imaging from \cite{Borlaff2019}. 

A blended source detection catalog is created by using \texttt{Photutils} \texttt{detect\_sources} on the SNR detection image with a threshold of SNR$\geq$1.8 over 10 pixels. This blended source detection catalog is then further deblended using \texttt{Photutils} \texttt{deblend\_sources} using the parameters nlevels = 10 and contrast =0.2 to isolate satellites of bright sources.


\subsection{Photometry} \label{Photometry}

The final segmentation map produced in Section \ref{Source_Detection} is used to construct our photometric catalog.  
Photometry is then preformed on our detected sources using Source Extraction and Photometry (SEP) \citep{ Barbary2016,Bertin96}. The source centriods are are determined by Source Extractors windowed algorithm. For each source, the SNR "detection" image is used to define the Gaussian-equivalent semi-major and semi-minor elliptical sizes, the elliptical orientation on the sky, the Kron radius, and the source centroids. Kron fluxes are then calculated with these parameters with Kron parameter of 2.5.
Our photometric uncertainties are measured by placing apertures with the same parameters as the those used to measure the source randomly across the source-masked mosaics, this allows our uncertainties to capture all sources of noise 
\citep{labb05,whit11,Whitaker2019}.
Aperture corrections are applied to the fluxes and uncertainties based on
Point Spread Functions (PSFs) for each band. 

\begin{figure}[tbp]
\centering
\epsscale{1.1}
\plotone{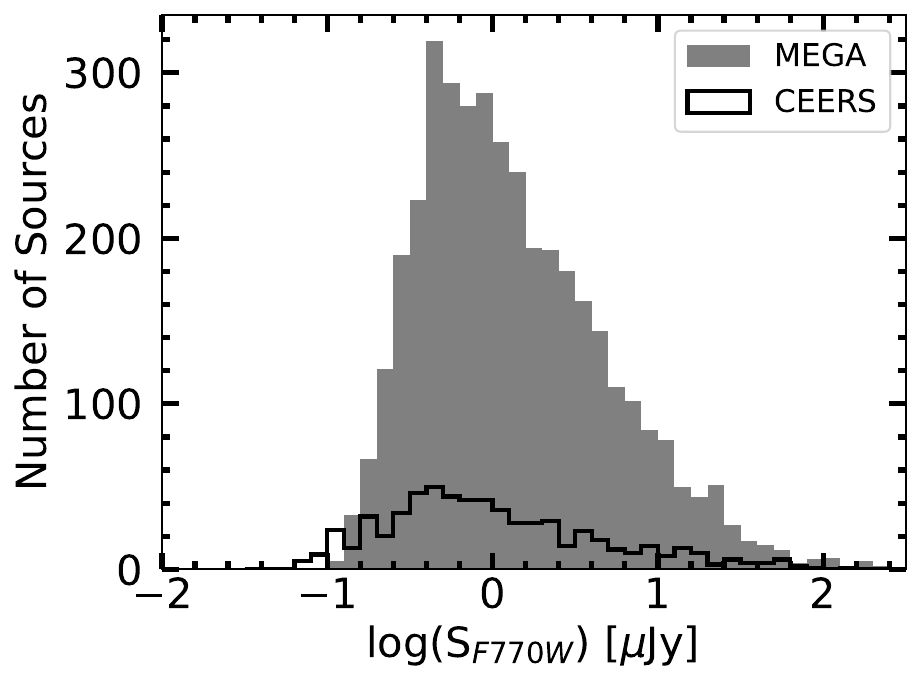}
\caption{Distribution of F770W photometric fluxes for our MEGA sample compared to CEERS. 
\label{fig:sources}} 
\end{figure}

Finally, We apply a signal-to-noise ratio (SNR) greater than 3 cut on the F770W sources. This reduced our final catalog from 5956 to 4444 detected sources. Figure \ref{fig:sources} shows the distribution of $3\sigma$ detections in MEGA F770W and $3\sigma$ detections in the CEERS MIRI observations. For this comparison the CEERS MIRI observations were re-reduced using the method outlined in Section \ref{Data_Reduction}.

\subsubsection{Quality Check}


False positive detections is evaluated
by comparing our source detection method on the re-reduced CEERS MIRI pointings to the CEERS MIRI catalog for the overlapping regions. We have 1057 sources with a 3$\sigma$ detection in at least one filter, 214 sources do not have a CEERS counterpart. Of these 214 galaxies 26.2\% are real sources. Of these real new sources 51.8\% are new multi-filter detected sources near the edge these detection differences are more likely due to potentially masking the Pointing edges. Additionally, 21.4\% of the new sources detections were multi-filter detections, and 26.8\% are single filter detections in the lowest filter. Of the 214 galaxies with no match 53.3\% are mischaracterized
extended features of brighter galaxies or noise features at the map edges. This makes our true false positive rate 3.9\%. Of these false positives, 22.2\% are in F770W, 36.1\% are in F1000W, 41.6\% are in F1500W, and 0\% in F2100W. Their appears to be two potential reasons for the higher false positive rate. Our super background methods makes use of the large number of observations, however due to the CEERS MIRI observation being taken over two separate periods the super background reductions typically only contain $\sim$8 \_cal.fits files whereas MEGA reductions (except for Pointing 18) had a super background made of $\sim$ 88\_cal.fits files. 

When we reduce our MEGA MIRI files using the CEERS reduction method and create a NIRCam prior catalog using TPHOT \citep{Merlin2015} similar to those done in \cite{Yang2023}. Once we select our regions that overlap with NIRCam, we have 2861 galaxies have a match. We additionally find 334 galaxies in our catalog which are not in the NIRCam TPHOT prior catalog. Of these 334 galaxies 41.9\% of these sources additional real sources, while 36.8\% are mischaracterized
extended features of brighter galaxies or noise features at the map edges. Similar to our CEERS MIRI quality check a majority of the additional real sources are multi-filter detections near the edge of the Pointing. This makes the final true false positive rate of 2.1\%. In both cases the false positive fluxes are at least a third of the lower than our true detections.

\section{Completeness and Number Counts}\label{Completeness_Num_Count}

\subsection{Image Depth}

We measure the 5$\sigma$ depth for our bands using the “ImageDepth” function of \texttt{PHOTUTILS}, which randomly places 100 circular nonoverlapping apertures on each mosaic avoiding detected sources. The 1$\sigma$ depth of an image is equal to the standard deviation of the Gaussian distribution of the resultant empty aperture fluxes. Therefore the 5$\sigma$ limiting flux are this standard deviation multiplied by the significance level (5). We set the radius of these circular apertures to be equal to the FWHM of each filter: 0.269'' (F770W), 0 .28'' (F1000W), 0.488'' (F1500W), and 0.674'' (F2100W). This 5$\sigma$ limiting flux has a filter-depended correction applied to obtain the total flux limit. The filter-depended correction is determined by WEBBPSF's \citep{Perrin2015} encircled energy fraction (EEF) versus radius curve. These 5$\sigma$ depths are shown in Table \ref{tab:filter_info}. 

\subsection{Number Counts}

We estimate the completeness of our source detection by using the Monte Carlo approach. We inject fake sources at different flux levels and note the fraction which are recovered using the same method of source detection and flux measurement applied to the real sources. For the injected fake sources, we model of their on the-sky distribution by stacking isolated and visually compact sources for each filter. 
We use compact sources because faint
galaxies are predominantly intrinsically small, and we do not find extended faint sources.

\begin{figure}[t!]
\centering
\epsscale{1.1}
\plotone{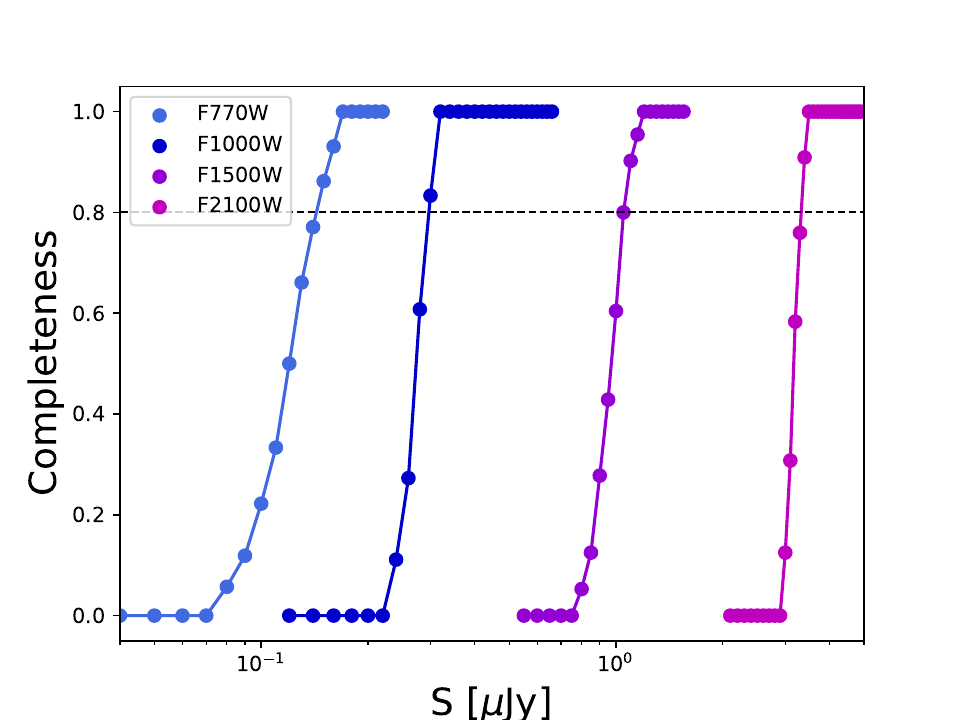}
\caption{Completeness fraction as a function of flux for for each of the MEGA MIRI filters. 80\% completeness for the F770W, F1000W, F1500W, and F2100W are 0.15, 0.30, 1.015, and 3.314 respectively. These 80\% completeness measurements  are comparable to the 5$\sigma$ Image Depth.
\label{fig:comp} }
\end{figure}

We then proceed to inject fake sources across a range of flux densities at random points in the image. At each flux, we injected $\sim$ 20 sources into the image by using the source model scaled to the desired flux density. We then process this image through the same process as that done for our real data. The completeness is the ratio of the number of recovered injected sources over the number of sources injected at each flux level. This process was done 100 times resulting in totaling 2,000 artificial sources per filter per flux bin and allowing to test the codes deblending ability better. The median completeness fraction is taken from all 100 runs. These completeness fractions are shown in Figure \ref{fig:comp}, along with noting 80\% completeness. We find an 80\% completeness of 0.14, 0.30, 1.08,and 3.31 $\mu$Jy for the F770W, F1000W, F1500W, and F2100W filters respectively. This 80\% completeness is comparable to the 5$\sigma$ limit shown in Table \ref{tab:filter_info}.

\begin{figure*}
\centering
\epsscale{1}
\plotone{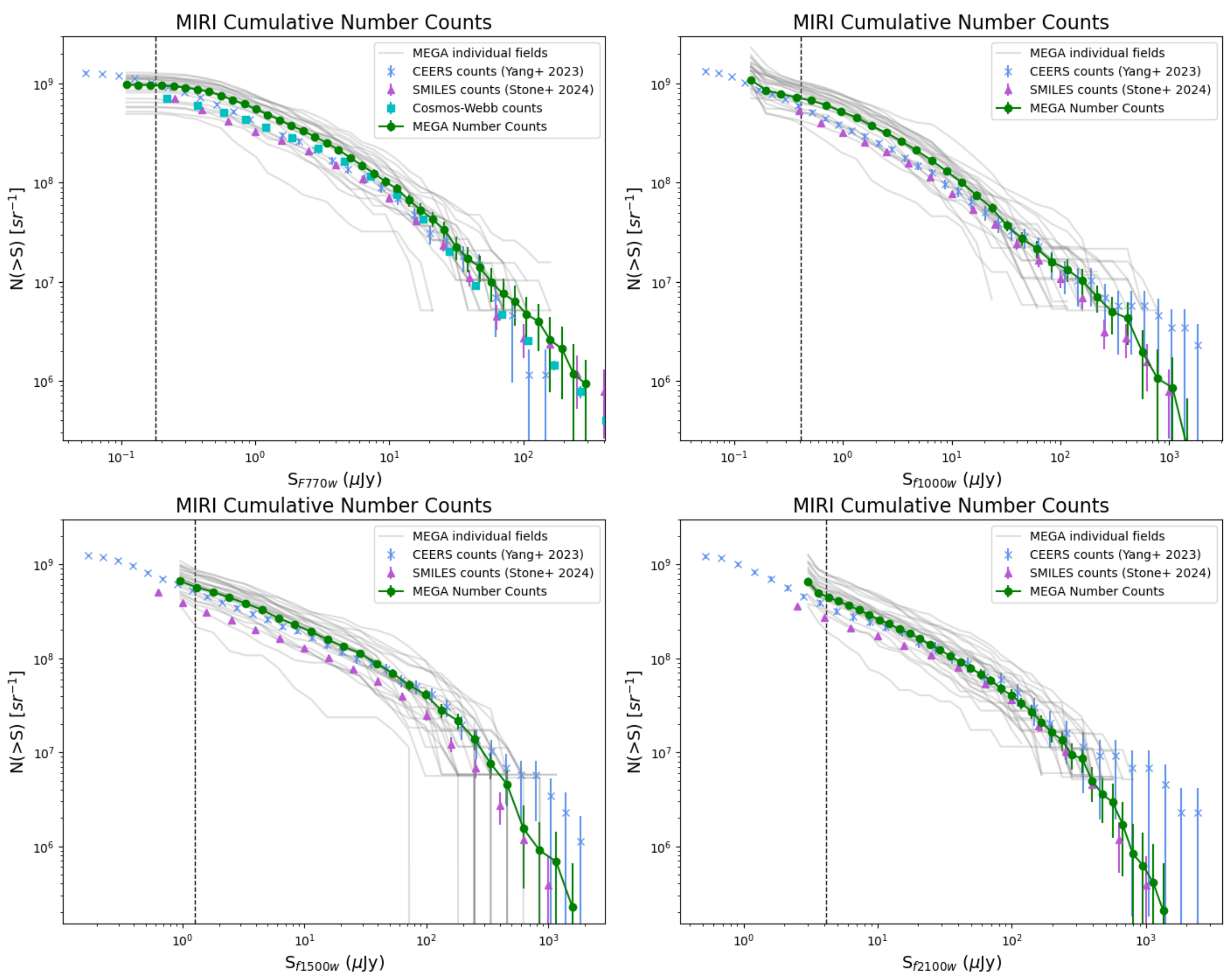}
\caption{The combined integral number counts for MEGA's four MIRI filters, with individual pointings shown in gray and the number counts using all pointings shown in green. The number counts of the total field have error bars representing the Poisson uncertainties calculated using ASTROPY.STATS.
POISSON\_CONF\_INTERVAL. As comparison the CEERs number counts from \cite{Yang2023} are shown as blue crosses, while SMILES number counts from \cite{Stone2024} are purple triangles. The dashed vertical line represents the 5$\sigma$ image depth for the MEGA pointings, which are shown in Table \ref{tab:filter_info}.
\label{fig:Num_count} }
\end{figure*}

The completeness corrected cumulative number counts are shown in Figure \ref{fig:Num_count}, where the uncertainties are estimated using \texttt{ASTROPY.STATS.POISSON\_CONF\_INTERVAL}. We followed the same method of obtaining our corrected cumulative number counts as \cite{sajkov2024}. While computing our number counts, we require a keep our initial SNR$>$3 F770W cutoff but require a SNR$>$1 for the other filters. Due to our observations being fairly uniform we assume the effective area does not change with flux level.

For comparison Figure \ref{fig:Num_count} shows our 5$\sigma$ limit for each filter along with CEERS, SMILES, and COSMOS-Web number counts \citep{Yang2023,Stone2024}. We find that our number counts are closer to SMILES at higher flux value in each individual filter while being closer to CEERS at lower fluxes. Some of our individual pointing number counts are consistent with those in the CEERS and SMILES number counts. These individual pointing number counts also show how much variation occurs between different fields. We note that the CEERS number counts from \cite{Yang2023} appear to go much deeper than our MEGA counts. However, the CEERS photometric catalog used NIRCam priors and after applying a SNR$>$3 cut on the CEERS F770W filter and SNR$>$1 cut on the other filters to match our catalog we get the similar minimum flux bins. This could be one of the reasons we see differences between the surveys number counts. Another potential cause of the differences between survey's is that both CEERS and SMILES use combined F560W and F770W images for source detection, while this work is done with the lowest MIRI filter available F770W or F1000W.

\section{Science Enabled by MEGA}\label{Science}

The primary science goals of the MEGA observational design are focused on the cosmic noon epoch ($0.5<z<3.5$) and are as follows:

\begin{itemize}
    \item Measure the obscured and unobscured star formation rate density and black hole accretion rate density for galaxies. 
    \item Measure the distribution of obscured and unobscured star formation in $\Ms$ galaxies.
    \item Identify and measure accretion rates of AGN in $M_\ast$ galaxies.

\end{itemize}

\subsection{Co-Evolution of Black Holes and Stars}
Previous knowledge of obscured star formation at $z>0.5$ has been historically limited to the brightest, dustiest galaxies due to the sensitivity limitations of {\it Spitzer} \citep[, and references therein]{case14}, or it has been indirectly estimated using attenutation measured from rest-frame optical and UV observations \citep[e.g.,][]{Reddy2016,fang18}.
\textit{Herschel} observations showed that the average attenuation increases when going backwards in time to peak at $z$ = 1-2 at $A_{FUV}$ = 2-2:5 mag \citep{buat2011} An accurate measurement of obscured and unobscured star formation within individual galaxies is vital to understanding the buildup of mass over time.

Similarly, measurements of black hole growth have been limited to X-ray luminous sources or the brightest infrared sources \citep[e.g.]{Delvecchio2014,Peca2023}. The more complete sampling of the IR SED enabled by MEGA will allow multiple detection methods including with CIGALE \citep[e.g.][]{Yang2023} and machine learning (Hamblin et al.\ 2025, in prep). 

With our rich data set, we are creating new luminosity functions to measure the unobscured and obscured SFR densities, and we are calculating new black hole accretion rate densities (Backhaus et al.\ 2025, in prep). MEGA will enable an in-depth look at how the SFR and black hole
accretion rate (BHAR) densities co-evolve, using the same population of galaxies for each calculation. This allows for a unique, direct comparison between the various densities. 

\subsection{The Morphology of Star Formation}
 MEGA observations allow us to reveal crucial gap by measuring the distribution of obscured star formation in $\Ms$ galaxies during the heyday of mass assembly.

By combining HST, NIRCam, and MIRI we are creating obscured SFR density ($\Sigma_{SFR,obscured}$), unobscured SFR density ($\Sigma_{SFR,unobscured}$), and stellar mass density($\Sigma_{\Ms}$) profiles for resolved galaxies. We are measuring the dust/star distributions as a function of $\Ms$, revealing any mismatch between dust and stellar morphology for moderately massed galaxies. Preliminary results indicate that galaxy morphology in the observed-frame mid-IR is significantly different from morphology in the near-IR, revealing more massive bulges and more merger signatures (Troiani et al.\, 2025, in prep\footnote{https://www.zooniverse.org/projects/gregtroiani/cosmic-collisions}).

\subsection{Finding AGN}


JWST MIRI's coverage allows us to find obscured and low-luminosity AGN, as the mid-IR is largely insensitive to obscuration. We are selecting AGN  in MEGA program by utilizing the existing 3-5 $\mu$m coverage from CEERS to observe stellar minimum in addition to the MEGA MIRI coverage to observe dust emission longwards of 6 $\mu$m (Kirkpatrick et al. 2025, in prep). Preliminary results show that at least 10\% of MEGA galaxies are classified as low-luminosity AGN at cosmic noon.
Additionally, our longer wavelength coverage is enabling the search for hot dust in little red dots, to confirm their AGN nature (Roynane et al.\,2025, Kocevski et al.\,2025)

\section{Summary}\label{Summary}

We present the observations, data reduction, photometric catalog, number counts, and science goals of the MIRI imaging from MEGA. MEGA is composed of 15 pointings covering 70 arcmin$^2$ of the EGS field using the F770W, F1000W, F1500W, and F2100W filters. Our reductions followed the method outlined in \cite{Alberts2024}, which uses the JWST Calibration Pipeline with additional steps to mask warm pixels, improving astrometry, removing detector artifacts, and improve the background subtraction. We show how multiple iterations of super background subtraction continue to improve the row and column stripping to create a more uniform background. We have extracted MIRI photometry using SEXTRACTOR. 

To test the quality of our source detection, we test for false positives two ways. We rereduce the CEERS obervations and use our detection method and fount a false positive rate of 3.5\%. However, we note our super background method makes use of a large set of observations and due to CEERS being split over two period the super backgrounds are created based on 8 \_cal.fits files which is 11 times smaller than our MEGA observations. Additionally, we rereduce our observations to match the CEERS reductions and detection method found a false positive rate of 2.1\%. Our photometic reductions also found 140 new source increasing the catalog by 4.6\%.

We have estimated the imaging depth for each pointing/band to be 0.18$\mu$m, 0.41$\mu$m, 1.26$\mu$m, and 4.1$\mu$m for the F770W, F1000W, F1500W, and F2100W filters respectively. These are comparable to 80\% completeness we derive. This completeness is then used to derive number counts for the fields. Our total number counts are comparable to CEERS and SMILES, and the deviation could be explained by difference in the fields, as shown by the deviation in our individual MIRI pointings number counts. These assessments indicate that our MIRI photometry catalog is of good quality and is able to complete our science goals.

\section{Acknowledgments}
B.E.B. and A.K. would like to especially thank Dr.\ Stacey Alberts for many helpful conversations about MIRI reductions. 
We acknowledge the hard work and support everyone involved in the JWST mission, especially Joy (Wilson) Skipper. B.E.B. and A.K. acknowledge support from grant JWST-GO-03794.001.
The MEGA JWST data presented in this paper were obtained from
the Mikulski Archive for Space Telescopes (MAST) at the Space Telescope Science Institute.

\software{\texttt{AstroPy} \citep{astropy2013,astropy2022}, \texttt{Matplotlib} \citep{hunter2007}, \texttt{NumPy} \citep{vanderwalt2011}, \texttt{SciPy} \citep{jones2001}, \texttt{eazy-py} \citep{bram08}, ASTRORMS, Photutils \citep{Bradley2020};JWST Calibration Pipeline \citep{bushouse_howard_2022_7429939}; SEXTRACTOR \citep{Bertin96};SEP \citep{Barbary2016}}

\bibliography{lib}{}

\begin{thebibliography}{}
\expandafter\ifx\csname natexlab\endcsname\relax\def\natexlab#1{#1}\fi
\providecommand{\url}[1]{\href{#1}{#1}}

\bibitem[{{Alberts} {et~al.}(2024){Alberts}, {Lyu}, {Shivaei}, {Rieke}, {P{\'e}rez-Gonz{\'a}lez}, {Bonaventura}, {Zhu}, {Helton}, {Ji}, {Morrison}, {Robertson}, {Stone}, {Sun}, {Williams}, \& {Willmer}}]{Alberts2024}
{Alberts}, S., {Lyu}, J., {Shivaei}, I., {et~al.} 2024, \apj, 976, 224

\bibitem[{{Astropy Collaboration} {et~al.}(2013){Astropy Collaboration}, {Robitaille}, {Tollerud}, {Greenfield}, {Droettboom}, {Bray}, {Aldcroft}, {Davis}, {Ginsburg}, {Price-Whelan}, {Kerzendorf}, {Conley}, {Crighton}, {Barbary}, {Muna}, {Ferguson}, {Grollier}, {Parikh}, {Nair}, {Unther}, {Deil}, {Woillez}, {Conseil}, {Kramer}, {Turner}, {Singer}, {Fox}, {Weaver}, {Zabalza}, {Edwards}, {Azalee Bostroem}, {Burke}, {Casey}, {Crawford}, {Dencheva}, {Ely}, {Jenness}, {Labrie}, {Lim}, {Pierfederici}, {Pontzen}, {Ptak}, {Refsdal}, {Servillat}, \& {Streicher}}]{astropy2013}
{Astropy Collaboration}, {Robitaille}, T.~P., {Tollerud}, E.~J., {et~al.} 2013, \aap, 558, A33

\bibitem[{{Astropy Collaboration} {et~al.}(2022){Astropy Collaboration}, {Price-Whelan}, {Lim}, {Earl}, {Starkman}, {Bradley}, {Shupe}, {Patil}, {Corrales}, {Brasseur}, {N{\"o}the}, {Donath}, {Tollerud}, {Morris}, {Ginsburg}, {Vaher}, {Weaver}, {Tocknell}, {Jamieson}, {van Kerkwijk}, {Robitaille}, {Merry}, {Bachetti}, {G{\"u}nther}, {Aldcroft}, {Alvarado-Montes}, {Archibald}, {B{\'o}di}, {Bapat}, {Barentsen}, {Baz{\'a}n}, {Biswas}, {Boquien}, {Burke}, {Cara}, {Cara}, {Conroy}, {Conseil}, {Craig}, {Cross}, {Cruz}, {D'Eugenio}, {Dencheva}, {Devillepoix}, {Dietrich}, {Eigenbrot}, {Erben}, {Ferreira}, {Foreman-Mackey}, {Fox}, {Freij}, {Garg}, {Geda}, {Glattly}, {Gondhalekar}, {Gordon}, {Grant}, {Greenfield}, {Groener}, {Guest}, {Gurovich}, {Handberg}, {Hart}, {Hatfield-Dodds}, {Homeier}, {Hosseinzadeh}, {Jenness}, {Jones}, {Joseph}, {Kalmbach}, {Karamehmetoglu}, {Ka{\l}uszy{\'n}ski}, {Kelley}, {Kern}, {Kerzendorf}, {Koch}, {Kulumani}, {Lee}, {Ly}, {Ma}, {MacBride}, {Maljaars}, {Muna}, {Murphy}, {Norman},
  {O'Steen}, {Oman}, {Pacifici}, {Pascual}, {Pascual-Granado}, {Patil}, {Perren}, {Pickering}, {Rastogi}, {Roulston}, {Ryan}, {Rykoff}, {Sabater}, {Sakurikar}, {Salgado}, {Sanghi}, {Saunders}, {Savchenko}, {Schwardt}, {Seifert-Eckert}, {Shih}, {Jain}, {Shukla}, {Sick}, {Simpson}, {Singanamalla}, {Singer}, {Singhal}, {Sinha}, {Sip{\H{o}}cz}, {Spitler}, {Stansby}, {Streicher}, {{\v{S}}umak}, {Swinbank}, {Taranu}, {Tewary}, {Tremblay}, {de Val-Borro}, {Van Kooten}, {Vasovi{\'c}}, {Verma}, {de Miranda Cardoso}, {Williams}, {Wilson}, {Winkel}, {Wood-Vasey}, {Xue}, {Yoachim}, {Zhang}, {Zonca}, \& {Astropy Project Contributors}}]{astropy2022}
{Astropy Collaboration}, {Price-Whelan}, A.~M., {Lim}, P.~L., {et~al.} 2022, \apj, 935, 167

\bibitem[{{Bagley} {et~al.}(2023){Bagley}, {Finkelstein}, {Koekemoer}, {Ferguson}, {Arrabal Haro}, {Dickinson}, {Kartaltepe}, {Papovich}, {P{\'e}rez-Gonz{\'a}lez}, {Pirzkal}, {Somerville}, {Willmer}, {Yang}, {Yung}, {Fontana}, {Grazian}, {Grogin}, {Hirschmann}, {Kewley}, {Kirkpatrick}, {Kocevski}, {Lotz}, {Medrano}, {Morales}, {Pentericci}, {Ravindranath}, {Trump}, {Wilkins}, {Calabr{\`o}}, {Cooper}, {Costantin}, {de la Vega}, {Hilbert}, {Hutchison}, {Larson}, {Lucas}, {McGrath}, {Ryan}, {Wang}, \& {Wuyts}}]{Bagley2023}
{Bagley}, M.~B., {Finkelstein}, S.~L., {Koekemoer}, A.~M., {et~al.} 2023, \apjl, 946, L12

\bibitem[{{Barbary}(2016)}]{Barbary2016}
{Barbary}, K. 2016, The Journal of Open Source Software, 1, 58

\bibitem[{{Bertin} \& {Arnouts}(1996)}]{Bertin96}
{Bertin}, E., \& {Arnouts}, S. 1996, \aaps, 117, 393

\bibitem[{{Borlaff} {et~al.}(2019){Borlaff}, {Trujillo}, {Rom{\'a}n}, {Beckman}, {Eliche-Moral}, {Infante-S{\'a}inz}, {Lumbreras-Calle}, {de Almagro}, {G{\'o}mez-Guijarro}, {Cebri{\'a}n}, {Dorta}, {Cardiel}, {Akhlaghi}, \& {Mart{\'\i}nez-Lombilla}}]{Borlaff2019}
{Borlaff}, A., {Trujillo}, I., {Rom{\'a}n}, J., {et~al.} 2019, \aap, 621, A133

\bibitem[{{Bouchet} {et~al.}(2015){Bouchet}, {Garc{\'\i}a-Mar{\'\i}n}, {Lagage}, {Amiaux}, {Augu{\'e}res}, {Bauwens}, {Blommaert}, {Chen}, {Detre}, {Dicken}, {Dubreuil}, {Galdemard}, {Gastaud}, {Glasse}, {Gordon}, {Gougnaud}, {Guillard}, {Justtanont}, {Krause}, {Leboeuf}, {Longval}, {Martin}, {Mazy}, {Moreau}, {Olofsson}, {Ray}, {Rees}, {Renotte}, {Ressler}, {Ronayette}, {Salasca}, {Scheithauer}, {Sykes}, {Thelen}, {Wells}, {Wright}, \& {Wright}}]{Bouchet2015}
{Bouchet}, P., {Garc{\'\i}a-Mar{\'\i}n}, M., {Lagage}, P.~O., {et~al.} 2015, \pasp, 127, 612

\bibitem[{{Bradley} {et~al.}(2020){Bradley}, {Sip{\H{o}}cz}, {Robitaille}, {Tollerud}, {Vin{\'\i}cius}, {Deil}, {Barbary}, {Wilson}, {Busko}, {G{\"u}nther}, {Cara}, {Conseil}, {Bostroem}, {Droettboom}, {Bray}, {Andersen Bratholm}, {Lim}, {Barentsen}, {Craig}, {Pascual}, {Perren}, {Greco}, {Donath}, {De Val-Borro}, {Kerzendorf}, {Bach}, {Weaver}, {D'Eugenio}, {Souchereau}, \& {Ferreira}}]{Bradley2020}
{Bradley}, L., {Sip{\H{o}}cz}, B., {Robitaille}, T., {et~al.} 2020, {astropy/photutils: 1.0.0}, v1.0.0,  Zenodo, doi:10.5281/zenodo.4044744

\bibitem[{{Brammer} {et~al.}(2008){Brammer}, {van Dokkum}, \& {Coppi}}]{bram08}
{Brammer}, G.~B., {van Dokkum}, P.~G., \& {Coppi}, P. 2008, \apj, 686, 1503

\bibitem[{{Buat} {et~al.}(2011){Buat}, {Giovannoli}, {Heinis}, {Charmandaris}, {Coia}, {Daddi}, {Dickinson}, {Elbaz}, {Hwang}, {Morrison}, {Dasyra}, {Aussel}, {Altieri}, {Dannerbauer}, {Kartaltepe}, {Leiton}, {Magdis}, {Magnelli}, \& {Popesso}}]{buat2011}
{Buat}, V., {Giovannoli}, E., {Heinis}, S., {et~al.} 2011, \aap, 533, A93

\bibitem[{Bushouse {et~al.}(2022)Bushouse, Eisenhamer, Dencheva, Davies, Greenfield, Morrison, Hodge, Simon, Grumm, Droettboom, Slavich, Sosey, Pauly, Miller, Jedrzejewski, Hack, Davis, Crawford, Law, Gordon, Regan, Cara, MacDonald, Bradley, Shanahan, Jamieson, Teodoro, \& Williams}]{bushouse_howard_2022_7429939}
Bushouse, H., Eisenhamer, J., Dencheva, N., {et~al.} 2022, JWST Calibration Pipeline, v1.8.5,  Zenodo, {If you use this software in your work, please cite it using the following metadata.}, doi:10.5281/zenodo.7429939.
\newblock \url{https://doi.org/10.5281/zenodo.7429939}

\bibitem[{{Casey} {et~al.}(2014){Casey}, {Narayanan}, \& {Cooray}}]{case14}
{Casey}, C.~M., {Narayanan}, D., \& {Cooray}, A. 2014, \physrep, 541, 45

\bibitem[{{Delvecchio} {et~al.}(2014){Delvecchio}, {Gruppioni}, {Pozzi}, {Berta}, {Zamorani}, {Cimatti}, {Lutz}, {Scott}, {Vignali}, {Cresci}, {Feltre}, {Cooray}, {Vaccari}, {Fritz}, {Le Floc'h}, {Magnelli}, {Popesso}, {Oliver}, {Bock}, {Carollo}, {Contini}, {Le F{\'e}vre}, {Lilly}, {Mainieri}, {Renzini}, \& {Scodeggio}}]{Delvecchio2014}
{Delvecchio}, I., {Gruppioni}, C., {Pozzi}, F., {et~al.} 2014, \mnras, 439, 2736

\bibitem[{{Fang} {et~al.}(2018){Fang}, {Faber}, {Koo}, {Rodr{\'\i}guez-Puebla}, {Guo}, {Barro}, {Behroozi}, {Brammer}, {Chen}, {Dekel}, {Ferguson}, {Gawiser}, {Giavalisco}, {Kartaltepe}, {Kocevski}, {Koekemoer}, {McGrath}, {McIntosh}, {Newman}, {Pacifici}, {Pandya}, {P{\'e}rez-Gonz{\'a}lez}, {Primack}, {Salmon}, {Trump}, {Weiner}, {Willner}, {Acquaviva}, {Dahlen}, {Finkelstein}, {Finlator}, {Fontana}, {Galametz}, {Grogin}, {Gruetzbauch}, {Johnson}, {Mobasher}, {Papovich}, {Pforr}, {Salvato}, {Santini}, {van der Wel}, {Wiklind}, \& {Wuyts}}]{fang18}
{Fang}, J.~J., {Faber}, S.~M., {Koo}, D.~C., {et~al.} 2018, \apj, 858, 100

\bibitem[{{Fazio} {et~al.}(2004){Fazio}, {Ashby}, {Barmby}, {Hora}, {Huang}, {Pahre}, {Wang}, {Willner}, {Arendt}, {Moseley}, {Brodwin}, {Eisenhardt}, {Stern}, {Tollestrup}, \& {Wright}}]{Fazio2004}
{Fazio}, G.~G., {Ashby}, M.~L.~N., {Barmby}, P., {et~al.} 2004, \apjs, 154, 39

\bibitem[{{Finkelstein} {et~al.}(2025){Finkelstein}, {Bagley}, {Arrabal Haro}, {Dickinson}, {Ferguson}, {Kartaltepe}, {Kocevski}, {Koekemoer}, {Lotz}, {Papovich}, {Perez-Gonzalez}, {Pirzkal}, {Somerville}, {Trump}, {Yang}, {Yung}, {Fontana}, {Grazian}, {Grogin}, {Kewley}, {Kirkpatrick}, {Larson}, {Pentericci}, {Ravindranath}, {Wilkins}, {Almaini}, {Amorin}, {Barro}, {Bhatawdekar}, {Bisigello}, {Brooks}, {Buitrago}, {Calabro}, {Castellano}, {Cheng}, {Cleri}, {Cole}, {Cooper}, {Cooper}, {Costantin}, {Cox}, {Croton}, {Daddi}, {Davis}, {Dekel}, {Elbaz}, {Fernandez}, {Fujimoto}, {Gandolfi}, {Gardner}, {Gawiser}, {Giavalisco}, {Gomez-Guijarro}, {Guo}, {Gupta}, {Hathi}, {Harish}, {Henry}, {Hirschmann}, {Hu}, {Hutchison}, {Iyer}, {Jaskot}, {Jha}, {Jung}, {Kokorev}, {Kurczynski}, {Leung}, {Llerena}, {Long}, {Lucas}, {Lu}, {McGrath}, {McIntosh}, {Merlin}, {Morales}, {Napolitano}, {Pacucci}, {Pandya}, {Rafelski}, {Rodighiero}, {Rose}, {Santini}, {Seille}, {Simons}, {Shen}, {Straughn}, {Tacchella}, {Vanderhoof},
  {Vega-Ferrero}, {Weiner}, {Willmer}, {Zhu}, {Bell}, {Wuyts}, {Holwerda}, {Wang}, {Wang}, \& {Zavala}}]{Finkelstein2025}
{Finkelstein}, S.~L., {Bagley}, M.~B., {Arrabal Haro}, P., {et~al.} 2025, arXiv e-prints, arXiv:2501.04085

\bibitem[{{Gardner} {et~al.}(2023){Gardner}, {Mather}, {Abbott}, {Abell}, {Abernathy}, {Abney}, {Abraham}, {Abraham}, {Abul-Huda}, {Acton}, {Adams}, {Adams}, {Adler}, {Adriaensen}, {Aguilar}, {Ahmed}, {Ahmed}, {Ahmed}, {Albat}, {Albert}, {Alberts}, {Aldridge}, {Allen}, {Allen}, {Altenburg}, {Altunc}, {Alvarez}, {{\'A}lvarez-M{\'a}rquez}, {Alves de Oliveira}, {Ambrose}, {Anandakrishnan}, {Andersen}, {Anderson}, {Anderson}, {Anderson}, {Anderson}, {Aprea}, {Archer}, {Arenberg}, {Argyriou}, {Arribas}, {Artigau}, {Arvai}, {Atcheson}, {Atkinson}, {Averbukh}, {Aymergen}, {Bacinski}, {Baggett}, {Bagnasco}, {Baker}, {Balzano}, {Banks}, {Baran}, {Barker}, {Barrett}, {Barringer}, {Barto}, {Bast}, {Baudoz}, {Baum}, {Beatty}, {Beaulieu}, {Bechtold}, {Beck}, {Beddard}, {Beichman}, {Bellagama}, {Bely}, {Berger}, {Bergeron}, {Bernier}, {Bertch}, {Beskow}, {Betz}, {Biagetti}, {Birkmann}, {Bjorklund}, {Blackwood}, {Blazek}, {Blossfeld}, {Bluth}, {Boccaletti}, {Boegner}, {Bohlin}, {Boia}, {B{\"o}ker}, {Bonaventura}, {Bond},
  {Bosley}, {Boucarut}, {Bouchet}, {Bouwman}, {Bower}, {Bowers}, {Bowers}, {Boyce}, {Boyer}, {Boyer}, {Boyer}, {Boyer}, {Bradley}, {Brady}, {Brandl}, {Brannen}, {Breda}, {Bremmer}, {Brennan}, {Bresnahan}, {Bright}, {Broiles}, {Bromenschenkel}, {Brooks}, {Brooks}, {Brown}, {Brown}, {Brown}, {Bruce}, {Bryson}, {Bujanda}, {Bullock}, {Bunker}, {Bureo}, {Burt}, {Bush}, {Bushouse}, {Bussman}, {Cabaud}, {Cale}, {Calhoon}, {Calvani}, {Canipe}, {Caputo}, {Cara}, {Carey}, {Case}, {Cesari}, {Cetorelli}, {Chance}, {Chandler}, {Chaney}, {Chapman}, {Charlot}, {Chayer}, {Cheezum}, {Chen}, {Chen}, {Cherinka}, {Chichester}, {Chilton}, {Chittiraibalan}, {Clampin}, {Clark}, {Clark}, {Clark}, {Claybrooks}, {Cleveland}, {Cohen}, {Cohen}, {Col{\'o}n}, {Coleman}, {Colina}, {Comber}, {Comeau}, {Comer}, {Conde Reis}, {Connolly}, {Conroy}, {Contos}, {Contreras}, {Cook}, {Cooper}, {Cooper}, {Correia}, {Correnti}, {Cossou}, {Costanza}, {Coulais}, {Cox}, {Coyle}, {Cracraft}, {Crew}, {Curtis}, {Cusveller}, {Da Costa Maciel}, {Dailey},
  {Daugeron}, {Davidson}, {Davies}, {Davis}, {Davis}, {Day}, {de Chambure}, {de Jong}, {De Marchi}, {Dean}, {Decker}, {Delisa}, {Dell}, \& {Dellagatta}}]{Gardner2023}
{Gardner}, J.~P., {Mather}, J.~C., {Abbott}, R., {et~al.} 2023, \pasp, 135, 068001

\bibitem[{{Grogin} {et~al.}(2011){Grogin}, {Kocevski}, {Faber}, {Ferguson}, {Koekemoer}, {Riess}, {Acquaviva}, {Alexander}, {Almaini}, {Ashby}, {Barden}, {Bell}, {Bournaud}, {Brown}, {Caputi}, {Casertano}, {Cassata}, {Castellano}, {Challis}, {Chary}, {Cheung}, {Cirasuolo}, {Conselice}, {Roshan Cooray}, {Croton}, {Daddi}, {Dahlen}, {Dav{\'e}}, {de Mello}, {Dekel}, {Dickinson}, {Dolch}, {Donley}, {Dunlop}, {Dutton}, {Elbaz}, {Fazio}, {Filippenko}, {Finkelstein}, {Fontana}, {Gardner}, {Garnavich}, {Gawiser}, {Giavalisco}, {Grazian}, {Guo}, {Hathi}, {H{\"a}ussler}, {Hopkins}, {Huang}, {Huang}, {Jha}, {Kartaltepe}, {Kirshner}, {Koo}, {Lai}, {Lee}, {Li}, {Lotz}, {Lucas}, {Madau}, {McCarthy}, {McGrath}, {McIntosh}, {McLure}, {Mobasher}, {Moustakas}, {Mozena}, {Nandra}, {Newman}, {Niemi}, {Noeske}, {Papovich}, {Pentericci}, {Pope}, {Primack}, {Rajan}, {Ravindranath}, {Reddy}, {Renzini}, {Rix}, {Robaina}, {Rodney}, {Rosario}, {Rosati}, {Salimbeni}, {Scarlata}, {Siana}, {Simard}, {Smidt}, {Somerville}, {Spinrad},
  {Straughn}, {Strolger}, {Telford}, {Teplitz}, {Trump}, {van der Wel}, {Villforth}, {Wechsler}, {Weiner}, {Wiklind}, {Wild}, {Wilson}, {Wuyts}, {Yan}, \& {Yun}}]{grog11}
{Grogin}, N.~A., {Kocevski}, D.~D., {Faber}, S.~M., {et~al.} 2011, \apjs, 197, 35

\bibitem[{{Hickox} \& {Alexander}(2018)}]{Hickox2018}
{Hickox}, R.~C., \& {Alexander}, D.~M. 2018, \araa, 56, 625

\bibitem[{{Hodge} {et~al.}(2015){Hodge}, {Riechers}, {Decarli}, {Walter}, {Carilli}, {Daddi}, \& {Dannerbauer}}]{Hodge2015}
{Hodge}, J.~A., {Riechers}, D., {Decarli}, R., {et~al.} 2015, \apjl, 798, L18

\bibitem[{{Hodge} {et~al.}(2016){Hodge}, {Swinbank}, {Simpson}, {Smail}, {Walter}, {Alexander}, {Bertoldi}, {Biggs}, {Brandt}, {Chapman}, {Chen}, {Coppin}, {Cox}, {Dannerbauer}, {Edge}, {Greve}, {Ivison}, {Karim}, {Knudsen}, {Menten}, {Rix}, {Schinnerer}, {Wardlow}, {Weiss}, \& {van der Werf}}]{Hodge2016}
{Hodge}, J.~A., {Swinbank}, A.~M., {Simpson}, J.~M., {et~al.} 2016, \apj, 833, 103

\bibitem[{{Hunter}(2007)}]{hunter2007}
{Hunter}, J.~D. 2007, Computing in Science and Engineering, 9, 90

\bibitem[{{Jones} {et~al.}(2001){Jones}, {Oliphant}, {Peterson}, \& Others}]{jones2001}
{Jones}, E., {Oliphant}, T., {Peterson}, P., \& Others. 2001, {SciPy}: Open Source Scientific Tools for Python, , .
\newblock \url{http://www.scipy.org/}

\bibitem[{{Kim} {et~al.}(2024){Kim}, {Goto}, {Ling}, {Wu}, {Hashimoto}, {Kilerci}, {Ho}, {Uno}, {Wang}, \& {Lin}}]{Kim2024}
{Kim}, S.~J., {Goto}, T., {Ling}, C.-T., {et~al.} 2024, \mnras, 527, 5525

\bibitem[{{Kirkpatrick} {et~al.}(2023){Kirkpatrick}, {Yang}, {Le Bail}, {Troiani}, {Bell}, {Cleri}, {Elbaz}, {Finkelstein}, {Hathi}, {Hirschmann}, {Holwerda}, {Kocevski}, {Lucas}, {McKinney}, {Papovich}, {P{\'e}rez-Gonz{\'a}lez}, {de la Vega}, {Bagley}, {Daddi}, {Dickinson}, {Ferguson}, {Fontana}, {Grazian}, {Grogin}, {Arrabal Haro}, {Kartaltepe}, {Kewley}, {Koekemoer}, {Lotz}, {Pentericci}, {Pirzkal}, {Ravindranath}, {Somerville}, {Trump}, {Wilkins}, \& {Yung}}]{kirkpatrick2023}
{Kirkpatrick}, A., {Yang}, G., {Le Bail}, A., {et~al.} 2023, \apjl, 959, L7

\bibitem[{{Koekemoer} {et~al.}(2011){Koekemoer}, {Faber}, {Ferguson}, {Grogin}, {Kocevski}, {Koo}, {Lai}, {Lotz}, {Lucas}, {McGrath}, {Ogaz}, {Rajan}, {Riess}, {Rodney}, {Strolger}, {Casertano}, {Castellano}, {Dahlen}, {Dickinson}, {Dolch}, {Fontana}, {Giavalisco}, {Grazian}, {Guo}, {Hathi}, {Huang}, {van der Wel}, {Yan}, {Acquaviva}, {Alexander}, {Almaini}, {Ashby}, {Barden}, {Bell}, {Bournaud}, {Brown}, {Caputi}, {Cassata}, {Challis}, {Chary}, {Cheung}, {Cirasuolo}, {Conselice}, {Roshan Cooray}, {Croton}, {Daddi}, {Dav{\'e}}, {de Mello}, {de Ravel}, {Dekel}, {Donley}, {Dunlop}, {Dutton}, {Elbaz}, {Fazio}, {Filippenko}, {Finkelstein}, {Frazer}, {Gardner}, {Garnavich}, {Gawiser}, {Gruetzbauch}, {Hartley}, {H{\"a}ussler}, {Herrington}, {Hopkins}, {Huang}, {Jha}, {Johnson}, {Kartaltepe}, {Khostovan}, {Kirshner}, {Lani}, {Lee}, {Li}, {Madau}, {McCarthy}, {McIntosh}, {McLure}, {McPartland}, {Mobasher}, {Moreira}, {Mortlock}, {Moustakas}, {Mozena}, {Nandra}, {Newman}, {Nielsen}, {Niemi}, {Noeske}, {Papovich},
  {Pentericci}, {Pope}, {Primack}, {Ravindranath}, {Reddy}, {Renzini}, {Rix}, {Robaina}, {Rosario}, {Rosati}, {Salimbeni}, {Scarlata}, {Siana}, {Simard}, {Smidt}, {Snyder}, {Somerville}, {Spinrad}, {Straughn}, {Telford}, {Teplitz}, {Trump}, {Vargas}, {Villforth}, {Wagner}, {Wandro}, {Wechsler}, {Weiner}, {Wiklind}, {Wild}, {Wilson}, {Wuyts}, \& {Yun}}]{koek11}
{Koekemoer}, A.~M., {Faber}, S.~M., {Ferguson}, H.~C., {et~al.} 2011, \apjs, 197, 36

\bibitem[{{Labb{\'e}} {et~al.}(2005){Labb{\'e}}, {Huang}, {Franx}, {Rudnick}, {Barmby}, {Daddi}, {van Dokkum}, {Fazio}, {Schreiber}, {Moorwood}, {Rix}, {R{\"o}ttgering}, {Trujillo}, \& {van der Werf}}]{labb05}
{Labb{\'e}}, I., {Huang}, J., {Franx}, M., {et~al.} 2005, \apjl, 624, L81

\bibitem[{{Madau} \& {Dickinson}(2014)}]{mada14}
{Madau}, P., \& {Dickinson}, M. 2014, \araa, 52, 415

\bibitem[{{Merlin} {et~al.}(2015){Merlin}, {Fontana}, {Ferguson}, {Dunlop}, {Elbaz}, {Bourne}, {Bruce}, {Buitrago}, {Castellano}, {Schreiber}, {Grazian}, {McLure}, {Okumura}, {Shu}, {Wang}, {Amor{\'\i}n}, {Boutsia}, {Cappelluti}, {Comastri}, {Derriere}, {Faber}, \& {Santini}}]{Merlin2015}
{Merlin}, E., {Fontana}, A., {Ferguson}, H.~C., {et~al.} 2015, \aap, 582, A15

\bibitem[{{Morrison} {et~al.}(2023){Morrison}, {Dicken}, {Argyriou}, {Ressler}, {Gordon}, {Regan}, {Cracraft}, {Rieke}, {Engesser}, {Alberts}, {Alvarez-Marquez}, {Colbert}, {Fox}, {Gasman}, {Law}, {Garcia Marin}, {G{\'a}sp{\'a}r}, {Guillard}, {Kendrew}, {Labiano}, {Laine}, {Noriega-Crespo}, {Shivaei}, \& {Sloan}}]{Morrison2023}
{Morrison}, J.~E., {Dicken}, D., {Argyriou}, I., {et~al.} 2023, \pasp, 135, 075004

\bibitem[{{Papovich} {et~al.}(2004){Papovich}, {Dole}, {Egami}, {Le Floc'h}, {P{\'e}rez-Gonz{\'a}lez}, {Alonso-Herrero}, {Bai}, {Beichman}, {Blaylock}, {Engelbracht}, {Gordon}, {Hines}, {Misselt}, {Morrison}, {Mould}, {Muzerolle}, {Neugebauer}, {Richards}, {Rieke}, {Rieke}, {Rigby}, {Su}, \& {Young}}]{Papovich2004}
{Papovich}, C., {Dole}, H., {Egami}, E., {et~al.} 2004, \apjs, 154, 70

\bibitem[{{Papovich} {et~al.}(2023){Papovich}, {Cole}, {Yang}, {Finkelstein}, {Barro}, {Buat}, {Burgarella}, {P{\'e}rez-Gonz{\'a}lez}, {Santini}, {Seill{\'e}}, {Shen}, {Arrabal Haro}, {Bagley}, {Bell}, {Bisigello}, {Calabr{\`o}}, {Casey}, {Castellano}, {Chworowsky}, {Cleri}, {Costantin}, {Cooper}, {Dickinson}, {Ferguson}, {Fontana}, {Giavalisco}, {Grazian}, {Grogin}, {Hathi}, {Holwerda}, {Hutchison}, {Kartaltepe}, {Kewley}, {Kirkpatrick}, {Kocevski}, {Koekemoer}, {Larson}, {Long}, {Lucas}, {Pentericci}, {Pirzkal}, {Ravindranath}, {Somerville}, {Trump}, {Urbano Stawinski}, {Weiner}, {Wilkins}, {Yung}, \& {Zavala}}]{Papovich2023}
{Papovich}, C., {Cole}, J.~W., {Yang}, G., {et~al.} 2023, \apjl, 949, L18

\bibitem[{{Peca} {et~al.}(2023){Peca}, {Cappelluti}, {Urry}, {LaMassa}, {Marchesi}, {Ananna}, {Balokovi{\'c}}, {Sanders}, {Auge}, {Treister}, {Powell}, {Turner}, {Kirkpatrick}, \& {Tian}}]{Peca2023}
{Peca}, A., {Cappelluti}, N., {Urry}, C.~M., {et~al.} 2023, \apj, 943, 162

\bibitem[{{P{\'e}rez-Gonz{\'a}lez} {et~al.}(2024){P{\'e}rez-Gonz{\'a}lez}, {Barro}, {Rieke}, {Lyu}, {Rieke}, {Alberts}, {Williams}, {Hainline}, {Sun}, {Pusk{\'a}s}, {Annunziatella}, {Baker}, {Bunker}, {Egami}, {Ji}, {Johnson}, {Robertson}, {Rodr{\'\i}guez Del Pino}, {Rujopakarn}, {Shivaei}, {Tacchella}, {Willmer}, \& {Willott}}]{Perez2024}
{P{\'e}rez-Gonz{\'a}lez}, P.~G., {Barro}, G., {Rieke}, G.~H., {et~al.} 2024, \apj, 968, 4

\bibitem[{{Perrin} {et~al.}(2015){Perrin}, {Long}, {Sivaramakrishnan}, {Lajoie}, {Elliot}, {Pueyo}, \& {Albert}}]{Perrin2015}
{Perrin}, M.~D., {Long}, J., {Sivaramakrishnan}, A., {et~al.} 2015, {WebbPSF: James Webb Space Telescope PSF Simulation Tool}, Astrophysics Source Code Library, record ascl:1504.007, ,

\bibitem[{{Reddy} {et~al.}(2016){Reddy}, {Steidel}, {Pettini}, \& {Bogosavljevi{\'c}}}]{Reddy2016}
{Reddy}, N.~A., {Steidel}, C.~C., {Pettini}, M., \& {Bogosavljevi{\'c}}, M. 2016, \apj, 828, 107

\bibitem[{{Rieke} {et~al.}(2015){Rieke}, {Wright}, {B{\"o}ker}, {Bouwman}, {Colina}, {Glasse}, {Gordon}, {Greene}, {G{\"u}del}, {Henning}, {Justtanont}, {Lagage}, {Meixner}, {N{\o}rgaard-Nielsen}, {Ray}, {Ressler}, {van Dishoeck}, \& {Waelkens}}]{Rieke2015}
{Rieke}, G.~H., {Wright}, G.~S., {B{\"o}ker}, T., {et~al.} 2015, \pasp, 127, 584

\bibitem[{{Rigby} {et~al.}(2023){Rigby}, {Perrin}, {McElwain}, {Kimble}, {Friedman}, {Lallo}, {Doyon}, {Feinberg}, {Ferruit}, {Glasse}, {Rieke}, {Rieke}, {Wright}, {Willott}, {Colon}, {Milam}, {Neff}, {Stark}, {Valenti}, {Abell}, {Abney}, {Abul-Huda}, {Acton}, {Adams}, {Adler}, {Aguilar}, {Ahmed}, {Albert}, {Alberts}, {Aldridge}, {Allen}, {Altenburg}, {{\'A}lvarez-M{\'a}rquez}, {Alves de Oliveira}, {Andersen}, {Anderson}, {Anderson}, {Argyriou}, {Armstrong}, {Arribas}, {Artigau}, {Arvai}, {Atkinson}, {Bacon}, {Bair}, {Banks}, {Barrientes}, {Barringer}, {Bartosik}, {Bast}, {Baudoz}, {Beatty}, {Bechtold}, {Beck}, {Bergeron}, {Bergkoetter}, {Bhatawdekar}, {Birkmann}, {Blazek}, {Blome}, {Boccaletti}, {B{\"o}ker}, {Boia}, {Bonaventura}, {Bond}, {Bosley}, {Boucarut}, {Bourque}, {Bouwman}, {Bower}, {Bowers}, {Boyer}, {Bradley}, {Brady}, {Braun}, {Breda}, {Bresnahan}, {Bright}, {Britt}, {Bromenschenkel}, {Brooks}, {Brooks}, {Brown}, {Brown}, {Brown}, {Bunker}, {Burger}, {Bushouse}, {Cale}, {Cameron}, {Cameron},
  {Canipe}, {Caplinger}, {Caputo}, {Cara}, {Carey}, {Carniani}, {Carrasquilla}, {Carruthers}, {Case}, {Catherine}, {Chance}, {Chapman}, {Charlot}, {Charlow}, {Chayer}, {Chen}, {Cherinka}, {Chichester}, {Chilton}, {Chonis}, {Clampin}, {Clark}, {Clark}, {Coe}, {Coleman}, {Comber}, {Comeau}, {Connolly}, {Cooper}, {Cooper}, {Coppock}, {Correnti}, {Cossou}, {Coulais}, {Coyle}, {Cracraft}, {Curti}, {Cuturic}, {Davis}, {Davis}, {Dean}, {DeLisa}, {deMeester}, {Dencheva}, {Dencheva}, {DePasquale}, {Deschenes}, {Hunor Detre}, {Diaz}, {Dicken}, {DiFelice}, {Dillman}, {Dixon}, {Doggett}, {Donaldson}, {Douglas}, {DuPrie}, {Dupuis}, {Durning}, {Easmin}, {Eck}, {Edeani}, {Egami}, {Ehrenwinkler}, {Eisenhamer}, {Eisenhower}, {Elie}, {Elliott}, {Elliott}, {Ellis}, {Engesser}, {Espinoza}, {Etienne}, {Etxaluze}, {Falini}, {Feeney}, {Ferry}, {Filippazzo}, {Fincham}, {Fix}, {Flagey}, {Florian}, {Flynn}, {Fontanella}, {Ford}, {Forshay}, {Fox}, {Franz}, {Fu}, {Fullerton}, {Galkin}, {Galyer}, {Garc{\'\i}a Mar{\'\i}n}, {Gardner},
  {Gardner}, {Garland}, {Garrett}, {Gasman}, {Gaspar}, {Gaudreau}, {Gauthier}, {Geers}, {Geithner}, {Gennaro}, {Giardino}, {Girard}, {Giuliano}, {Glassmire}, \& {Glauser}}]{Rigby2023}
{Rigby}, J., {Perrin}, M., {McElwain}, M., {et~al.} 2023, \pasp, 135, 048001

\bibitem[{{Ronayne} {et~al.}(2024){Ronayne}, {Papovich}, {Yang}, {Shen}, {Dickinson}, {Kennicutt}, {Alavi}, {Arrabal Haro}, {Bagley}, {Burgarella}, {Le Bail}, {Bell}, {Cleri}, {Cole}, {Costantin}, {de la Vega}, {Daddi}, {Elbaz}, {Finkelstein}, {Grogin}, {Holwerda}, {Kartaltepe}, {Kirkpatrick}, {Koekemoer}, {Lucas}, {Magnelli}, {Mobasher}, {P{\'e}rez-Gonz{\'a}lez}, {Prichard}, {Rafelski}, {Rodighiero}, {Sunnquist}, {Teplitz}, {Wang}, {Windhorst}, \& {Yung}}]{Ronayne2024}
{Ronayne}, K., {Papovich}, C., {Yang}, G., {et~al.} 2024, \apj, 970, 61

\bibitem[{{Sajkov} {et~al.}(2024){Sajkov}, {Sajina}, {Pope}, {Alberts}, {Armus}, {Farrah}, {Lin}, {Marchesini}, {McKinney}, {Veilleux}, {Yan}, \& {Young}}]{sajkov2024}
{Sajkov}, L., {Sajina}, A., {Pope}, A., {et~al.} 2024, \apj, 977, 115

\bibitem[{{Shipley} {et~al.}(2016){Shipley}, {Papovich}, {Rieke}, {Brown}, \& {Moustakas}}]{Shipley2016}
{Shipley}, H.~V., {Papovich}, C., {Rieke}, G.~H., {Brown}, M. J.~I., \& {Moustakas}, J. 2016, \apj, 818, 60

\bibitem[{{Spoon} {et~al.}(2007){Spoon}, {Marshall}, {Houck}, {Elitzur}, {Hao}, {Armus}, {Brandl}, \& {Charmandaris}}]{Spoon2007}
{Spoon}, H.~W.~W., {Marshall}, J.~A., {Houck}, J.~R., {et~al.} 2007, \apjl, 654, L49

\bibitem[{{Stone} {et~al.}(2024){Stone}, {Alberts}, {Rieke}, {Bunker}, {Lyu}, {P{\'e}rez-Gonz{\'a}lez}, {Shivaei}, \& {Zhu}}]{Stone2024}
{Stone}, M.~A., {Alberts}, S., {Rieke}, G.~H., {et~al.} 2024, \apj, 972, 62

\bibitem[{van~der Walt {et~al.}(2011)van~der Walt, Colbert, \& Varoquaux}]{vanderwalt2011}
van~der Walt, S., Colbert, S.~C., \& Varoquaux, G. 2011, Computing in Science \& Engineering, 13, 22.
\newblock \url{http://aip.scitation.org/doi/abs/10.1109/MCSE.2011.37}

\bibitem[{{Whitaker} {et~al.}(2011){Whitaker}, {Labb{\'e}}, {van Dokkum}, {Brammer}, {Kriek}, {Marchesini}, {Quadri}, {Franx}, {Muzzin}, {Williams}, {Bezanson}, {Illingworth}, {Lee}, {Lundgren}, {Nelson}, {Rudnick}, {Tal}, \& {Wake}}]{whit11}
{Whitaker}, K.~E., {Labb{\'e}}, I., {van Dokkum}, P.~G., {et~al.} 2011, \apj, 735, 86

\bibitem[{{Whitaker} {et~al.}(2019){Whitaker}, {Ashas}, {Illingworth}, {Magee}, {Leja}, {Oesch}, {van Dokkum}, {Mowla}, {Bouwens}, {Franx}, {Holden}, {Labb{\'e}}, {Rafelski}, {Teplitz}, \& {Gonzalez}}]{Whitaker2019}
{Whitaker}, K.~E., {Ashas}, M., {Illingworth}, G., {et~al.} 2019, \apjs, 244, 16

\bibitem[{{Wright} {et~al.}(2023){Wright}, {Rieke}, {Glasse}, {Ressler}, {Garc{\'\i}a Mar{\'\i}n}, {Aguilar}, {Alberts}, {{\'A}lvarez-M{\'a}rquez}, {Argyriou}, {Banks}, {Baudoz}, {Boccaletti}, {Bouchet}, {Bouwman}, {Brandl}, {Breda}, {Bright}, {Cale}, {Colina}, {Cossou}, {Coulais}, {Cracraft}, {De Meester}, {Dicken}, {Engesser}, {Etxaluze}, {Fox}, {Friedman}, {Fu}, {Gasman}, {G{\'a}sp{\'a}r}, {Gastaud}, {Geers}, {Glauser}, {Gordon}, {Greene}, {Greve}, {Grundy}, {G{\"u}del}, {Guillard}, {Haderlein}, {Hashimoto}, {Henning}, {Hines}, {Holler}, {Detre}, {Jahromi}, {James}, {Jones}, {Justtanont}, {Kavanagh}, {Kendrew}, {Klaassen}, {Krause}, {Labiano}, {Lagage}, {Lambros}, {Larson}, {Law}, {Lee}, {Libralato}, {Lorenzo Alverez}, {Meixner}, {Morrison}, {Mueller}, {Murray}, {Mycroft}, {Myers}, {Nayak}, {Naylor}, {Nickson}, {Noriega-Crespo}, {{\"O}stlin}, {O'Sullivan}, {Ottens}, {Patapis}, {Penanen}, {Pietraszkiewicz}, {Ray}, {Regan}, {Roteliuk}, {Royer}, {Samara-Ratna}, {Samuelson}, {Sargent}, {Scheithauer},
  {Schneider}, {Schreiber}, {Shaughnessy}, {Sheehan}, {Shivaei}, {Sloan}, {Tamas}, {Teague}, {Temim}, {Tikkanen}, {Tustain}, {van Dishoeck}, {Vandenbussche}, {Weilert}, {Whitehouse}, \& {Wolff}}]{Wright2023}
{Wright}, G.~S., {Rieke}, G.~H., {Glasse}, A., {et~al.} 2023, \pasp, 135, 048003

\bibitem[{{Yang} {et~al.}(2023){Yang}, {Papovich}, {Bagley}, {Ferguson}, {Finkelstein}, {Koekemoer}, {P{\'e}rez-Gonz{\'a}lez}, {Arrabal Haro}, {Bisigello}, {Caputi}, {Cheng}, {Costantin}, {Dickinson}, {Fontana}, {Gardner}, {Grazian}, {Grogin}, {Harish}, {Holwerda}, {Iani}, {Kartaltepe}, {Kewley}, {Kirkpatrick}, {Kocevski}, {Kokorev}, {Lotz}, {Lucas}, {Navarro-Carrera}, {Pentericci}, {Pirzkal}, {Ravindranath}, {Rinaldi}, {Shen}, {Somerville}, {Trump}, {de la Vega}, {Wilkins}, \& {Yung}}]{Yang2023}
{Yang}, G., {Papovich}, C., {Bagley}, M.~B., {et~al.} 2023, \apjl, 956, L12

\end{thebibliography}

\end{document}